\documentclass[aps,prb,twocolumn,superscriptaddress,longbibliography]{revtex4-2}

\usepackage{amsmath,amssymb}
\usepackage{bm}
\usepackage{graphicx}
\usepackage{epstopdf} 
\usepackage{latexsym} 
\usepackage{subfigure}
\usepackage{color}
\usepackage{dsfont} 
\usepackage{wasysym}

\newcommand{\mc}{\mathcal}

\begin{document}

\title{Infinite critical boson non-Fermi liquid}

\author{Xiao-Tian Zhang}
\author{Gang Chen}
\thanks{$^1$Department of Physics and HKU-UCAS Joint Institute for Theoretical and Computational Physics at Hong Kong, The University of Hong Kong, Pokfulam Road, Hong Kong SAR, China. 
$^2$The University of Hong Kong Shenzhen Institute of Research and Innovation, 518057, Shenzhen, Guangdong Province, China.  Email: \email{gangchen@hku.hk}}

\begin{abstract}
We study a distinct type of non-Fermi liquid where there exists an infinite 
number of critical bosonic modes instead of finite number of bosonic modes 
for the conventional ones. We consider itinerant magnets with both 
conduction electrons and fluctuating magnetic moments in three dimensions.  
With Dzyaloshinskii-Moriya interaction, the moments fluctuate 
near a boson surface in the reciprocal space at low energies when 
the system approaches an ordering transition. The infinite number  
of critical modes on the boson surface strongly scatter the  
gapless electrons on the Fermi surface and convert the metallic sector 
into a non-Fermi liquid. We explain the physical properties of this 
non-Fermi liquid. On the ordered side, a conventional non-Fermi liquid 
emerges due to the scattering by the gapless Goldstone mode from 
the spontaneous breaking of the global rotational symmetry. 
We discuss the general structure of the phase diagram in the vicinity 
of the quantum phase transition and clarify various crossover behaviors.
\end{abstract}

\maketitle

\noindent{\bf{\small Introduction.}}\\
Landau Fermi liquid theory is the major milestone of modern condensed 
matter physics, and illustrates the triumph of physical intuition~\cite{Landau}. 
The short-ranged repulsive interaction between the fermions was 
argued to be irrelevant as one approaches the low energy towards the Fermi surface. 
The singular long-range interactions, however, are not well coped in the framework 
of Fermi liquid theory and signifies the possibility of non-Fermi liquid (NFL)
metals~\cite{RMP2007,Hertz1976}. These singular interactions can 
come from (partially screened) long-range Coulomb interaction, 
the fluctuations of the gapless bosonic modes at the criticality, the
Goldstone boson from the continuous symmetry breaking~\cite{Watanabe_2014}, 
and the U(1) gauge boson~\cite{SSLee2018,SSLee2009}. The established
theories describing NFL metals, particularly the experimentally relevant 
ones, are known as Hertz-Millis-Moriya theory~\cite{Hertz1976,Millis1993,Moriya1973,Moriya1973B,Moriya2012}.
This theory involves the coupling between gapless fermions 
near the Fermi surface and the critical bosons. 
If the number of the gapless fermions is finite 
such as Dirac fermion, Weyl fermion and the quadratic band touching,
a controlled calculation with the perturbative renormalization group  
can be performed. In contrast, when the fermion sector is a Fermi surface, 
the physics become complex and this topic is under an active investigation
in recent years. 
 


\begin{figure}[b] 
	\centering
	\includegraphics[width=6.5cm]{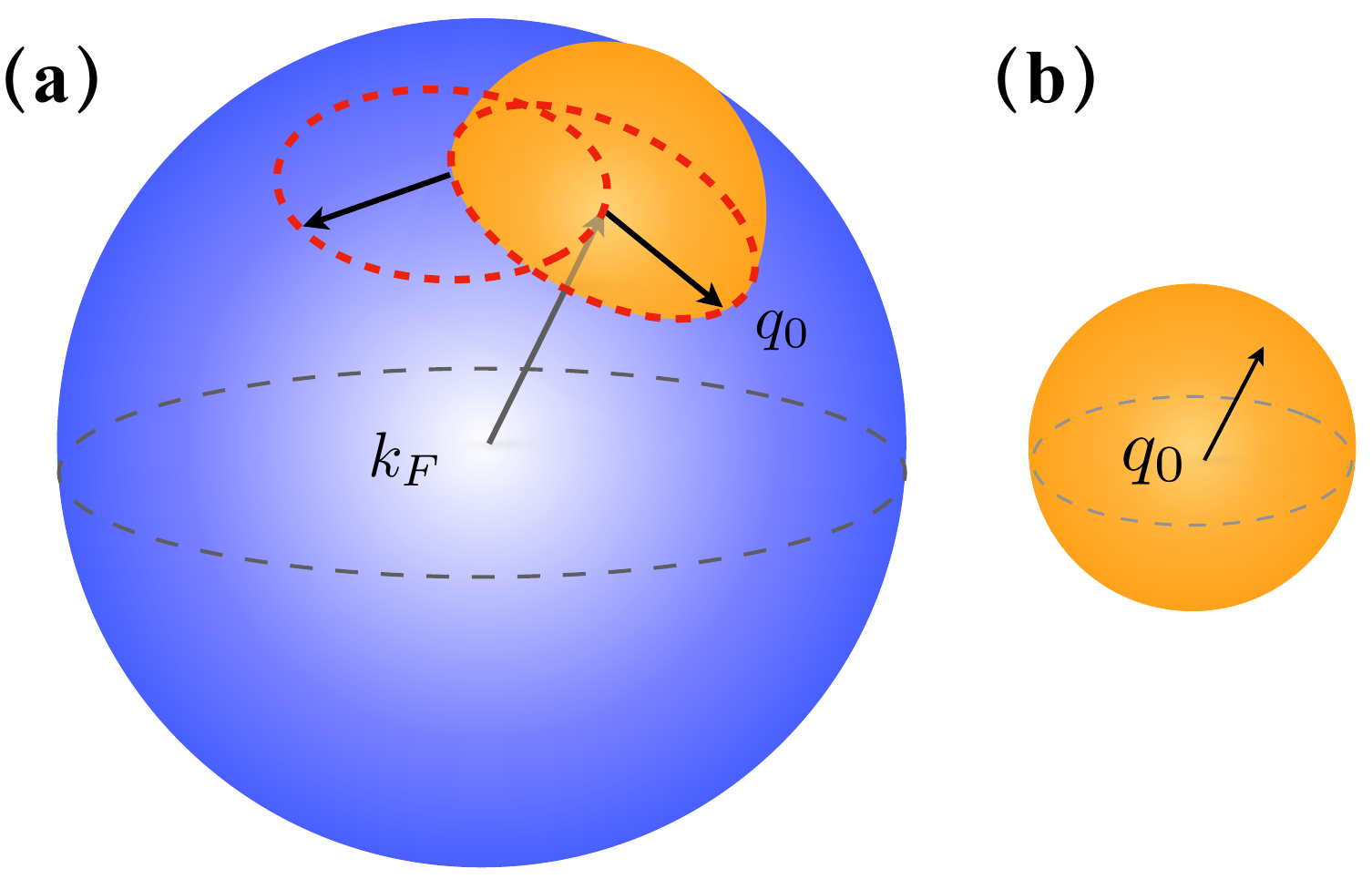}
	\caption{{\bf The coupling between 
	Fermi surface and critical boson surface.}
	(a) The blue sphere is the 
	Fermi surface of the conduction electrons.
	The orange surface represents
	the critical boson surface.
	$k_{\rm F}$ is the Fermi momentum, 
	and $q_0$ is the radius of the boson surface.
	The Yukawa coupling connects fermions 
	with a ring	of fermion modes on the Fermi surface.
	(b) The critical boson sphere. 
	}
	\label{fig1}
\end{figure}

So far there are two types of Fermi surface criticality associated NFLs~\cite{SSLee2018}. 
The first involves an ordering at a finite wavevector, e.g. an antiferromagnetic (AFM) order~\cite{Chubukov2000,Chubukov2004,Abrahams2012,SSLee2015,SSLee2017} 
or spin density wave~\cite{Sachdev2010B}. The ordering wavevector connects a few 
`hotspots' on the Fermi surface, and the theoretical analysis focuses on the coupling 
between the hotspot fermions and the critical bosons. The second involves the critical 
bosons at the zero wavevector. This captures, for instance, the Ising-nematic 
criticality~\cite{Fradkin2001,Metzner2003,Fradkin2007,Sachdev2010,SSLee2013}, 
spinon-gauge coupling in the spinon Fermi surface U(1) spin liquid~\cite{SSLee2009}, 
ferromagnetic (FM) criticality~\cite{Chubukov2006,Chubukov2009}. 
It was observed that, the bosons that are tangential to the Fermi surface 
scatter the fermions strongly at the low energies. Thus, the theoretical analysis 
of this Ising-nematic criticality is further reduced to the so-called patch theory 
where the tangential critical boson scatters the fermions from one patch or two patches. 
Various analytic techniques were developed. The early random phase approximation
type of large-$N$ expansion was questioned as this scheme of taming 
quantum fluctuations and organizing the Feynman diagrams misses the contribution 
from the processes involving the fermions on the Fermi surface~\cite{Polchinski1994,Millis1994,Kim1995,SSLee2009,Sachdev2010}. 
The remedy was made by the double expansion that combines the large-${\mathit N}$
expansion and the $\epsilon$-expansion~\cite{Nayak1994,Senthil2010}. 
Another remedy introduces the dimensional and co-dimensional 
regularization to the Fermi surface, and develops a systematic framework to 
regulate the quantum fluctuations~\cite{PRL1995,Shankar2009,SSLee2013,SSLee2015,SSLee2015B}. 
It is hoped that, the physical cases are located in the regimes 
where these development can be applied. Inspired by these   
developments, we turn our attention to another type of Fermi surface 
criticality and NFL. Compared to the efforts in the literature, here we are more 
inclined to exploring the mechanism and phenomenology of the 
NFLs.

In this paper, we study the system with the Fermi surface coupled to 
the critical bosons on a continuously degenerate manifold,
i.e. a boson surface in three dimensional space(3D) as shown   
in Fig.~\ref{fig1}. The critical phase of bosons in one dimension
is the Luttinger liquid~\cite{Giamarchi2003}.
Focusing on the stability of such highly degenerate critical phase in higher dimensions 
has been investigated in the context of 
Bose metal~\cite{Das1999,Phillips2003,Paramekanti2002,Motrunich2007,Sheng2009,Han2021},
and recent studies declared that weakly interacting dilute bosonic 
systems with continuously degenerate minima in the low-energy bosonic excitations 
are stable in ${\mathit d}=2,3$ and thereby pointed to the concept of Bose Luttinger liquid~\cite{Sur2019,Lake2021}. 
Experimentally, the relevance of critical boson surface has been implied 
by the neutron scattering in MnSi~\cite{Pfleiderer2001,Pfleiderer2004}
where a nearly uniform intensity is measured on a sphere. Motivated by these developments, 
the investigation on the critical boson surface has received some attentions~\cite{Tsvelik2021,Ku2021,Zhang2022}.
Here we are not dealing with nor relying on the stable phase of critical boson surfaces, 
instead we aim to improve our understanding of the critical boson surface 
induced quantum criticality and its impact when it is coupled with gapless fermions.

We consider the 3D itinerant magnets that comprise two distinct types of degrees of freedom:
i) conduction electrons, ii) local magnetic moments. In the absence of inversion symmetry, 
there exists a Dzyaloshinskii-Moriya interaction between the local moments that is responsible 
for the generation of infinite critical bosons on a spherical surface at the phase transition.
The critical boson surface is coupled to the fermions on the Fermi surface 
at the low energies through a Yukawa-type interaction. 
Microscopically, this Yukawa coupling arises from the Hund's or Kondo-like coupling
between the conduction electron and the local moments. 
As illustrated in Fig.~\ref{fig1}(a), 
each fermion is coupled to a ring of fermions on the Fermi surface, 
and the fermions on this ring are further coupled to many other rings of fermions. 
Thus, infinite number of gapless fermions are scattered 
by the infinite number of critical bosons, 
which makes the whole Fermi surface critical.
This fermion-boson-coupled model is fundamentally different
from the AFM criticality or the Ising-nematic criticality where 
only a finite number of critical bosons are involved. 
Thus, neither the hotspot treatment for the AFM criticality 
nor the conventional patch theory is applicable. 
We establish the basic properties and the global phase diagram of the system,
and adopt the self-consistent renormalization theory~\cite{Moriya2012} 
to address the properties near the criticality. We show that, 
due to the novel type of fermion-boson coupling, 
the system becomes a NFL metal with distinct power-law behaviors 
in the vicinity of the transition. We analyze the fermion and the boson 
properties as well as the related crossovers 
at the criticality and in the ordered regime. 
\\

\begin{figure}[t] 
	\centering
	\includegraphics[width=8.7cm]{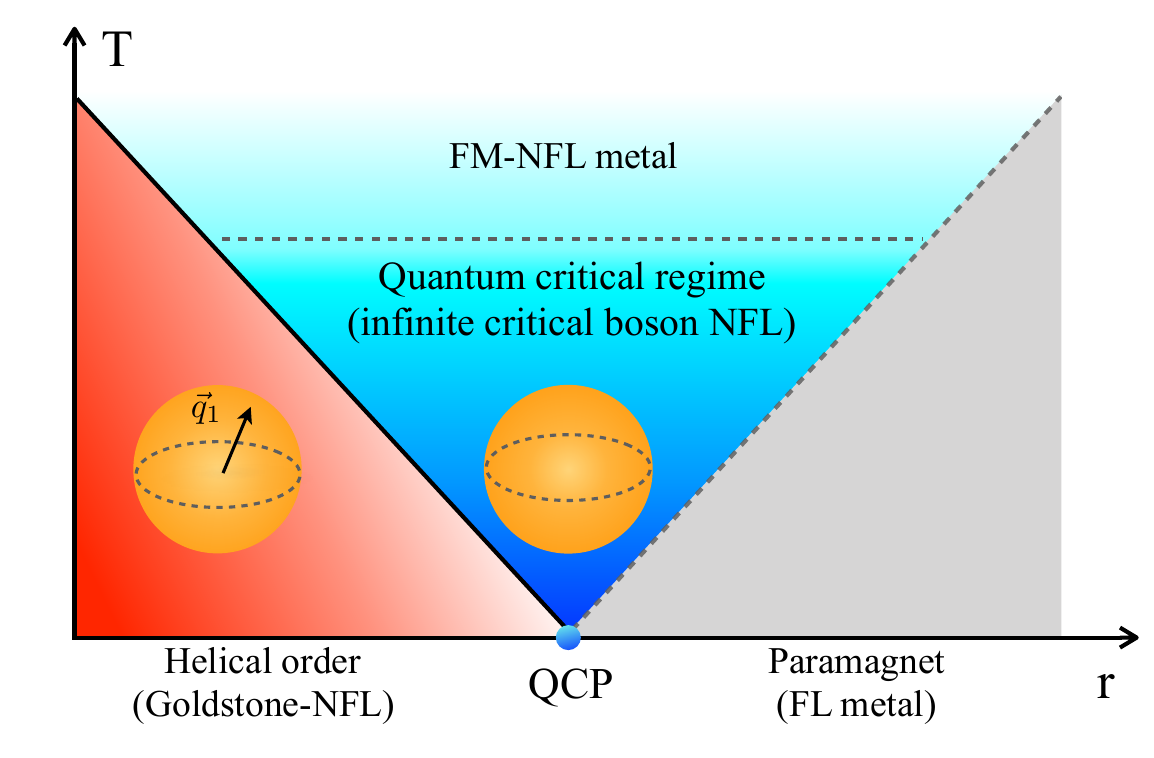}
	\caption{{\bf The global phase diagram.}
	The left corner is a helimagnet and a Goldstone-NFL where the NFL is induced by the gapless Goldstone boson. 
	The right corner is a paramagnet and Fermi liquid metal. The central region is the quantum critical regime where  
    a distinct NFL with infinite critical bosons on the boson surface is realized. 
    As the temperature rises to the point where the thermal fluctuation submerges the boson surface, 
    the system experiences a crossover to a FM criticality-like NFL behavior. 
    The solid (dashed) line refers to the phase transition (thermal crossover).
     	}
	\label{fig2}
\end{figure}

\noindent{\bf Results}\\
\noindent{\bf{\small Model.}} 
The full model for the itinerant magnets contains three parts, 
the conducting electrons, the local moments and the coupling between them,
which is then described by a fermion-boson-coupled Lagrangian
\begin{eqnarray}
 {\cal L}[f^\dagger,f;\vec{\phi}] & = & 
  \sum_{l,\alpha} f^\dagger_{l\alpha}(\partial_\tau -\mu) f_{l\alpha}^{} 
- \sum_{ll^\prime,\alpha} t_{ll^\prime}^{} f^\dagger_{l\alpha} f_{l^\prime\alpha}^{}
\nonumber 
\\
& + & g \sum_{l,\alpha\beta} f^\dagger_{l\alpha} \vec{\sigma}_{\alpha\beta}^{} 
               f_{l\beta}^{} \cdot \vec{\phi}_{l}^{}
  + {\cal L}_{\rm B}^{} [\vec{\phi}].
  \label{eq1}
\end{eqnarray}
The first line of Eq.~\eqref{eq1} dictates an electron model 
where $f^\dagger(f)$ is the fermion creation (annihilation) operator. 
The itinerant electrons hop on a 3D lattice denoted by the site index $l$. 
The electron spin couples to a magnetic moment field $\vec{\phi}_l$ 
with a Kondo-like Yukawa coupling. 
Here, the bosonic field $\vec{\phi}_l$ is a three-component
vector defined on site $l$, $\vec{\sigma}_{\alpha\beta}$ is the 
Pauli matrix vector with $\alpha, \beta$ being the spin indices. 
The magnetic fluctuation near the phase transition admits a standard
Ginzburg-Landau expansion up to $\mathcal{O}(\phi^4)$ with
\begin{equation}
{\cal L}_{\rm B}[\vec{\phi}] = \sum_{l} \frac{1}{2} \vec{\phi}_l \cdot (r-J\nabla^2) \vec{\phi}_l 
+ \frac{u}{4} (\vec{\phi}_l^2)^2 + \frac{D}{2} \vec{\phi}_l \cdot (\nabla \times \vec{\phi}_l) ,
\label{phi_4}
\end{equation}
The first two terms of ${\cal L}_{\rm B}[ \vec{\phi}]$ represent a standard $\phi^4$-theory 
with an order parameter $\vec{\phi}$ for the magnetic moment and $r$ is the boson mass.
The last term is an anti-symmetric Dzyaloshinskii-Moriya (DM) interaction, which fundamentally 
alters the critical phenomenon and leads to a rich phase diagram in Fig.~\ref{fig2}.

\begin{figure}[t] 
	\centering
	\includegraphics[width=8.5cm]{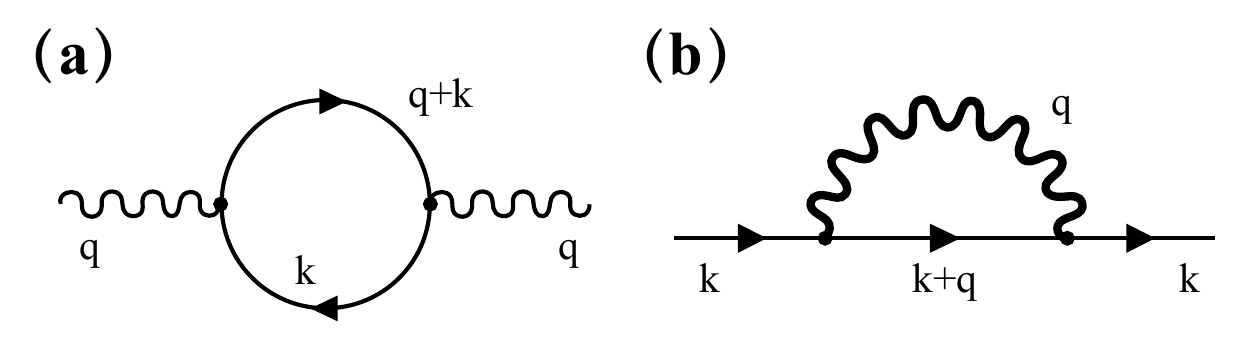}
	\caption{{\bf Renormalized bubble diagrams.}
	(a) The fermion bubble induced boson dynamics.  
	(b) The renormalized fermion propagator from the renormalized 
	boson correlator. The light and bold curly line represent the 
	bare and renormalized boson correlators, respectively. 
	}
	\label{fig3}
\end{figure}

To tackle with the bosonic fluctuation, a saddle point solution of $\vec{\phi}$ is 
required, on top of which the lowest order expansion counts for the fluctuations. 
As shown in Fig.~\ref{fig3}, we follow the Hertz approach by integrating out 
the gapless fermions. This gives rise to the Landau damping that dominates 
the low-energy boson dynamics. The effective action for the bosonic sector 
is written as,
\begin{equation}
\begin{aligned}
\mc{S}_{\rm B} = & 
\frac{1}{2}\sum_{{\bf q}, i\omega_l} {\it \Pi}_{\mu\nu}({\bf q},i\omega_l) 
\phi_{\mu}({\bf q}, i \omega_l) \phi_{\nu}(-{\bf q},-i\omega_l) \\
& + \frac{u}{4}\int d\tau\int d^3{\bf r} \big[\vec{\phi}^2({\bf r},\tau)\big]^2 ,\\
\end{aligned}
\label{S_B}
\end{equation}
where $\mu,\nu,\lambda$ label the vector components of $\vec{\phi}$, 
and we have converted the 3D lattice index to a continuous, 
real space coordinate ${\bf r}$. 
The fermionic bubble is illustrated in Fig.~\ref{fig3}
and the renormalized boson polarization takes the form 
\begin{equation}
\begin{aligned}
& {\it \Pi}_{\mu\nu}({\bf q},i\omega_l) = f(q,i\omega_l)\,  \delta_{\mu\nu} 
          - i  D \epsilon_{\mu\nu\lambda} \,q_{\lambda} , \\
& f(q,i\omega_l) = r+ J {q}^2 + \frac{|\omega_l|}{{\it \Gamma}_q}  ,\\
\end{aligned}
\label{boson_Pi_2}
\end{equation}
where  ${\omega_l=2{\mathrm \pi} l \beta^{-1}} $ ${(l\in {\mathbb Z})}$ 
is the Matsubara frequency for the bosons, and the $|\omega_l|/{\it \Gamma}_q$
is the Landau damping term. In general, the function 
${\it \Gamma}_q$ takes a form ${{\it \Gamma}_q= {\it \Gamma} q}$ with 
${q =|{\bf q}|}$. 
\\

\noindent{\bf{\small Critical boson surface.}} 
The DM interaction complicates the low-energy 
theories by introducing the vector index into the bosonic sector. 
The dispersion of the bosonic modes are modified 
compared to the ${D=0}$ case. Diagonalizing the bare quadratic bosonic 
part at ${\omega_{l}=0}$, we obtain three branches of bosonic modes 
with dispersions given by
\begin{equation}
{E}_{n}(q)=r+Jq^2+nDq, \quad n=-1,0,+1.
\label{E_n}
\end{equation}
The lowest branch ${E}_{-1}(q)$ is of particular interest, 
which reaches its minima on a spherical surface in the momentum space 
${q=q_0 \equiv D/(2J)}$. Approaching the criticality at ${r_{\rm c}=D^2/(4J)}$, 
the lowest mode ${E}_{-1}(q)$ becomes gapless at the surface ${q=q_0}$. Thus,
the boson modes on the entire sphere ${q=q_0}$ become critical simultaneously,
which is then dubbed `critical boson surface'. The finite radius of the sphere
is guaranteed by the DM interaction. The function ${{\it \Gamma}_{q}\approx {\it \Gamma} q_0}$,
and ${\it \Gamma}$ is a constant due to the finite density of states on the Fermi surface.

For a pure classical boson system, the large phase space provided by the critical boson surface 
always result in a fluctuation-driven first-order transition~\cite{Brazovskii1975}.
The low-energy dynamics of the fermion-boson coupled system 
is not determined by the critical boson surface alone;
rather the boson receives renormalization from the particle-hole excitation around Fermi surface.
A previous work by Schmalian and Turlakov~\cite{Schmalian2004}
shed light on the nature of the quantum phase transition in the presence of 
the critical boson surface for a fermion-boson coupled system.
The effective low-energy theory of the critical bosons are obtained 
by projecting onto the $n=-1$ mode in Eq.~\eqref{E_n},
which turns out to be a $\phi^4$-theory with multiple quartic interaction constants.
In certain parameter regimes, the transition can be a second order transition in the university class ${d=3, z=2}$
and a mean field theory of this second order transition has been developed~\cite{Vojta2001}.
In the following, we regard that our system undergoes a second-order quantum transition
and treat the critical fluctuation around the quantum critical point perturbatively.
This physical scenario is realized within a crossover regime ${\xi_{\rm Gi} \gg \xi \gg \xi_{\rm DM}}$~\cite{Schmalian2004,Janoschek2013}.
Here ${\xi_{\rm DM}\sim q_0^{-1}}$ is the length scale of DM interaction. 
When the correlation length ${\xi \gg \xi_{\rm DM}}$, 
the fluctuation of the bosons is dominated by the critical boson surface.
The opposite limit ${\xi \ll \xi_{\rm DM}}$ dictates a high temperature regime in the phase diagram of Fig.~\ref{fig2}.
In addition, we regard that the fluctuations are weakly interacting according to the Ginzburg criteria ${\xi \gg \xi_{\rm Gi}}$,
and will carry out the self-consistent renormalization study next.
\\

\noindent{\bf{\small Self-consistent renormalization theory.}} 
Although the patch and the hotspot theories are 
inadequate for our fermion-boson-coupled model,
a phenomenological technique, dubbed the self-consistent renormalization 
(SCR) theory~\cite{Moriya2012,nagaosa1999}, captures the key features of the fermion-boson 
coupling and provides the evidence for the behaviors near the criticality. 
The spirit of the SCR approach is to find the most appropriate quadratic 
action that encodes the effective renormalized non-linear interactions. 
This approach works well with $d=3, z=2$ and
quantitatively produces the experimental results in many itinerant magnets 
of various dimensions~\cite{RMP2007} and works for continuous and nearly continuous 
transitions~\cite{Moriya1973,Moriya1973B}. 
We here make an attempt to implement the SCR calculation for our fermion-boson-coupled 
model and hope to gain some understanding about the physical properties of the model.
Moreover, since the SCR approach has never been applied to the itinerant magnets 
with the degenerate low-energy critical modes, our attempt would add one physical example to the SCR theory.

To search for the best action in the SCR approach, one relies on Feynman's variational
method to optimize the free energy. Following the procedure by Moriya~\cite{Moriya2012},
we consider a trial quadratic action with the following 
form 
\begin{eqnarray}
\tilde{\mathcal S}_{\text B}[\tilde{r}] &=& \frac{1}{2}\sum_{{\bf q},i \omega_l} 
\Big[ \big( \tilde{r} + Jq^2+  \frac{|\omega_l|}{{\it \Gamma}_q}  \big) \delta_{\mu\nu} 
- i D\epsilon_{\mu\nu\lambda} q_{\lambda}
\Big]
\nonumber \\
&&\quad\quad\quad\quad \times  \phi_{\mu}({\bf q}, i \omega_l) \phi_{\nu}(-{\bf q},-i\omega_l)  ,
\label{S_B_var}
\end{eqnarray}
where we replace $r$ with a variational parameter $\tilde{r}$ that is to be determined. 
The boson correlator is given as a hermitian matrix $M_{\mu\nu}$ [see Supplementary Note 6 for derivations],
\begin{equation}
\langle \phi_{\mu} ({\bf q}, i \omega_l) \phi_{\nu} ({\bf q}', i \omega_{l'}) 
\rangle_{\tilde{\cal S}_{\rm B}}
\simeq
M_{\mu\nu} ({{\bf q}, i\omega_l})
\delta_{{\bf q} ,-{\bf q}'}\delta_{i \omega_l ,-i \omega_{l'}} 
\end{equation}
where $\langle \cdots \rangle_{\tilde{\cal S}_{\rm B}}$ refers 
to the statistical average against $\tilde{\cal S}_{\rm B}$.
This correlation conceives the information of the critical boson surface,
and serves as a key ingredient for the mechanism of the proposed NFL
that can be detected by the neutron scattering experiment. At the criticality,
the powder-averaged neutron scattering spectrum is given as 
$ \text{Tr} M({\bf q},\omega)\sim (\omega/{\it \Gamma}_{q_0})/\big[J^2(q-q_0)^{4}
+(\omega/{\it \Gamma}_{q_0})^2\big]$.
Around the critical boson surface ${|{\bf q}|=q_0}$, 
the spectrum displays a divergent behavior. In the real space, 
the boson surface momentum $q_0$ provides a characteristic length scale $1/q_0$,
which endows the correlation function with a spatial modulation in all directions,
and the correlator from the elastic neutron scattering is given by 
an envelop function on top of the usual power law decaying in the long-distance limit as
${\sum_{\mu} \langle \phi_{\mu}({\bf r}) \phi_{\mu}(0) \rangle_{\tilde{\cal S}_{\rm B}} \sim \sin(q_0 |{\bf r}|)/{ |{\bf r}|^2}}$.

 The variational free energy for the bosonic sector is 
\begin{eqnarray}
  F ({\tilde r}) &\equiv & \tilde{F} ({\tilde r}) + \frac{1}{\beta} \langle {\mathcal S}_{\text B} 
                    - \tilde{\mathcal S}_{\text B} \rangle_{\tilde{\cal S}_{\text B}} 
                    \nonumber \\
                    &=&   \tilde{F} ({\tilde r})
                    + 
                   \frac{ 1}{2\beta} ( r - \tilde{r}   )   \big[ \text{Tr }  
                          \sum_{ {\bf q}, i \omega_l   } M ({{\bf q}, i\omega_l}) \big] +
                          \frac{u}{4\beta^2 V} 
\nonumber \\                          
                      &\times &   
                     \Big\{ 
                   \big[ \text{Tr }   
                          \sum_{ {\bf q}, i \omega_l   } M ({{\bf q}, i\omega_l} ) 
                   \big]^2 
                          + 2 {\text{Tr}} \big[\sum_{{\bf q}, i \omega_l }
                         M^{}  ({{\bf q}, i\omega_l} )\big]^2
                   \Big\}, 
                   \nonumber \\                          
\end{eqnarray}
where $\tilde{F} ({\tilde r})$ is the free energy corresponding to the trial 
quadratic action $\tilde{\cal S}_{\text B}$, and $V$ is the system volume. 
The variational parameter $\tilde{r}$ is determined from the saddle point 
equation ${\partial_{\tilde{r}} F(\tilde{r}) =0}$. 
After the cumbersome calculation detailed in Supplementary Note 1,
the optimization procedure results in the following self-consistent equation
for the parameter ${\tilde{r}}$,
\begin{eqnarray}
\tilde{r} - r = \frac{uc}{\beta V} \sum_{n=\pm 1,0}
\sum_{{\bf q}, i \omega_l } \frac{1}{ |\omega_l |/{\it \Gamma}_q + {\cal E}_{n} (q)},
\label{SCR_r}
\end{eqnarray}
where ${{\cal E}_{n} (q)={E}_{n} (q)|_{r\rightarrow \tilde{r} }}$ and $c$ is a constant prefactor. 
We further set ${\delta \equiv \tilde{r} - r_{\rm c}}$ and ${\delta_0 \equiv r - r_{\rm c}}$. Here $\delta$ 
measures the distance from the quantum critical point ${\delta_c=0}$ and is related to 
the correlation length $\xi(T)$ that is expected to diverge at the criticality at low temperatures with ${\delta(T) =\xi^{-2}(T)}$. 
From the above self-consistent equation, we find that the quantum 
fluctuation at finite temperatures is encoded in the function $\delta(T)$ 
that scales as [see Supplementary Note 2 for derivations]
\begin{eqnarray}
\delta(T) \sim T^{\alpha} , \ \ \ \ {\alpha= \frac{4}{5}}.
\label{eqn_alpha}
\end{eqnarray}
It is illuminating to compare with the behaviors for the 3D FM (AFM) criticality
where the dynamic exponent ${z=3}$ (${z=2}$) and the SCR calculation yields 
${\alpha={4}/{3} }$ (${ \alpha={3}/{2} }$). In fact, the same exponents 
were obtained from a simple scaling counting using Millis's renormalization 
${\alpha=(d+z-2)/z}$~\cite{RMP2007}. For both FM and AFM criticality, 
${\alpha >1}$.  In contrast, ${\alpha<1}$ in Eq.~\eqref{eqn_alpha} indicates 
much stronger fluctuations due to the extensive phase space provided by the critical boson surface.
\\

\noindent{\bf{\small NFL behavior from critical boson surface.}} 
Unlike the FM or AFM criticalities where the low-energy fluctuations are at discrete momenta, 
the low-energy fluctuations near a finite boson surface strongly
scatter the itinerant electrons and reduce the lifetime of electron
quasiparticles. We use the renormalized boson correlator 
and the Feymann diagram in Fig.~\ref{fig3} to calculate the 
self-energy of the conduction electron, 
\begin{equation}
\begin{aligned}
\Sigma ( {\bf k}, i\epsilon_n ) \simeq 
\frac{g^2}{\beta V} \sum_{{\bf q},i\omega_l}
G_0({\bf k}+{\bf q},i\epsilon_n+i\omega_l) 
\text{Tr} [ M({\bf q},i\omega_l) ] ,
\end{aligned}
\end{equation}
where ${\epsilon_n\equiv(2n+1)\pi /\beta}$ ${(n \in {\mathbb Z})}$ is the fermion 
Matsubara frequency, and ${G_0({\bf k},i\epsilon_n)= (i\epsilon_n-\xi_{\bf k})^{-1}}$ 
is a bare Green's function of the electrons with a dispersion ${\xi_{\bf k} \equiv {{\bf k}^2}/{(2m)}-\epsilon_{\rm F}}$. 
By performing an analytic continuation ${i\epsilon_n \rightarrow \omega+i\eta}$,
we obtain the $T$-dependence for the retarded self-energy in the static limit ${\omega= 0}$ 
with ${|{\bf k}| =k_{\rm F}}$ located on the Fermi surface [see Supplementary Note 4 for derivations], 
\begin{equation}
{\rm Im} \Sigma^{\rm R}({\bf k},\omega=0;T) \sim  {T^3}{ \delta }^{-2} \sim T^{3-2\alpha}. 
\label{im_Sig}
\end{equation}
We note the $T$-dependence here has multiple sources that might render
a simple scaling $\sim \omega/T$ inappropriate. We can see this by evaluating 
the $\omega$-dependence of the self-energy in the zero temperature limit is evaluated in Supplementary Note 5,
which yields ${{\rm Im} \Sigma^{\rm R}({\bf k},\omega;T=0) \sim \sqrt{\omega}}$.
This result is consistent with literature~\cite{Schmalian2004}.

The imaginary part of the self-energy in Eq.~\eqref{im_Sig} determines the scattering 
rate for the forward scattering process provided the exchanged boson momentum is 
small ${q_0\ll k_{\rm F}}$. The electronic resistivity (or the inverse transport lifetime)
is obtained from the scattering rate by multiplying an angular factor ${(1-\cos\theta)}$,
where ${\theta \simeq q_0/k_{\rm F} \ll 1}$ is the small forward scattering angle.
The narrow scattering angle suppress the resistivity by ${\sim (q_0/k_{\rm F})^2}$,
and more importantly, the $T$-dependence is inherited from Eq.~\eqref{im_Sig} 
as,
\begin{equation}
\rho(T)\sim \frac{q_0^2}{2k_{\rm F}^2}{\rm Im} \Sigma^{\rm R}({\bf k},\omega=0;T) \sim T^{7/5}. 
\label{rho}
\end{equation}
This peculiar power-law $T$-dependence indicates a distinct NFL behavior owing to 
the scattering of the electrons by the extensive critical fluctuations on the boson surface.
The accuracy of the present study is limited by the phenomenological method we used,
namely we only considered small angle forward scattering. 
Whereas the large angle scattering processes enabled 
by multiple scatterings on the Fermi surface are neglected.
Moreover, the SCR method represents a way to bypass the significant challenge that
it is extremely difficult to cook up a low-energy effective theory.
\\

\noindent{\bf{\small Crossover to FM-NFL at high-$T$.}} We have discussed the NFL behavior arised
from the strong fluctuations near the boson surface at low-$T$ which is quoted as an infinite
critical boson NFL. When the temperature is further increased
to be larger than a characteristic energy associated 
with the boson sphere radius $q_0$, i.e. ${T\gg {\it \Gamma}_{q_0}}$,
the fluctuation is no longer dominated by the boson modes near the critical surface. 
In fact, the structure of the critical boson surface is no longer discernible at high temperatures.
The boson sphere can be regarded as a point object in the reciprocal space, 
resembling the case of the FM criticality. In this high temperature regime above the criticality,
the temperature dependence of the variational parameter crossovers to scale as
${\delta(T) \sim T^{{4}/{3}} }$, which coincides with the case of the NFL
from the FM fluctuations in 3D. The calculation is shown in Supplementary Note 3. Thus, the system undergoes 
a crossover between two distinct types of NFLs (see Fig.~\ref{fig2}), 
and the crossover temperature can be approximately set by the difference 
of the boson energies at the center and at the surface of the boson sphere, 
\begin{equation}
T_c \sim {E_{-1} (q=0) -E_{-1} (q_0) } \sim {\mathcal O} ( \frac {D^2}{4J}). 
\end{equation}
\\

\noindent{\bf{\small NFL in helimagnetic ordered phase.}} 
When ${r<r_{\rm c}}$, the system develops a magnetic
order by spontaneously selecting the ordering wavevector from the degenerate boson surface [see Supplementary Note 7 for more discussions]. 
In the left corner of phase diagram in Fig.~\ref{fig2}, a helical order with an ordering
wavevector at ${{\bf q}_1=q_{0} \hat{n}}$ is picked up where $\hat{n}$ is the propagating 
direction of the helimagnet. The original model in Eq.~\eqref{eq1} is invariant under
a combined rotation with respect to the real space and the internal space of the 
magnetic orders. The helimagnet spontaneously breaks this continuous symmetry 
and thus generates gapless Goldstone modes. In the helimagnetic phase, 
a small fluctuation above the helical order parameter couples to the itinerant electrons with 
\begin{eqnarray}
\sim g\int d^3{\bf r}\ f^\dagger_\alpha({\bf r}) 
\vec{\sigma}_{\alpha\beta} f_\beta({\bf r}) \cdot \delta \vec{\phi}({\bf r}),
\end{eqnarray}
where ${\delta \vec{\phi}({\bf r}) = \phi_0 [-\varphi({\bf r})\sin q_0z, \varphi({\bf r})\cos q_0z, \theta({\bf r}) ]} $
with $\theta({\bf r})$ and $\varphi({\bf r})$ describing the polar and the azimuthal phase fluctuations 
against the helical order ${\langle \vec{\phi} ({\bf r}) \rangle = \phi_0 [ \cos (q_0 z) \hat{x} + \sin (q_0 z) \hat{y} ]}$
with ${\hat{n}=\hat{z}}$.
These phase fluctuations give rise to gapless Goldstone modes~\cite{BK2006}.
The Yukawa coupling, $g$, remains finite in the low-energy limit. 
This is due to the fact that the generator for the continuous symmetry 
involves the orbital angular momentum and thus does not commute
with the translation operator, i.e. the total momentum. 
Applying the general criteria in Ref.~\onlinecite{Watanabe_2014},
we see that the Goldstone mode converts the fermion sector
into a NFL which is dubbed `Goldstone-NFL'. The NFL behavior in the helimagnetic phase, 
{\rm e.g.} the electronic resistivity, has been discussed\cite{BK2005,BK2006B}.
\\

\noindent{\bf Discussion.}\\
In summary, we have studied a distinct type of NFL in the
3D itinerant magnets that are not captured by the conventional patch and 
hotspot theories. The infinite critical boson modes on the boson sphere
connect all momentum points on the Fermi surface for the itinerant electrons,
leading to unconventional consequences for both local moments and itinerant electrons. 
The SCR approach adopted in our study is a phenomenological method
that has been proven to be successful in explaining the FM and the AFM 
fermion criticalities~\cite{Moriya2012}. For the continuously degenerate boson surface criticality,
the SCR approach incorporates the large scattering phase space between fermions and bosons
where the renormalized fermionic and bosonic correlations are taken into consideration sequentially.

On the more experimental side, several further investigations with material-based 
simulations can be expected. In reality, the boson sector would experience a cubic anisotropy 
in the interaction that would favor the wavevectors along either 001 or 111 directions and thus 
lift the degeneracy of the boson surface. This cubic anisotropy could set another crossover energy 
scale for the problem. The effect of external fields such as the pressure and magnetic field is anticipated  
due to its relevance to the experiments on MnSi~\cite{Pfleiderer2001,Pfleiderer2004}.
An anomalous NFL transport behavior is observed in a widespread region 
of the phase diagram spanned by pressure and magnetic field.
The experimental relevance of the infinite critical boson NFL to the physical systems
like MnSi may be addressed in a specific study with a more realistic consideration.
In general, the role of a finite uniform magnetic field is two-folded. 
Firstly, it leads to the procession of the spin order parameter,
which competes with the Landau damping dynamics caused by the fermion.
Secondly, this external field introduces anisotropy for the quantum critical points,
which may effectively reduce the fluctuating dimension of the boson sector.
Intuitively, the 2D counterpart of the present problem can be readily considered
where a Fermi circle is coupled to a 1D critical boson contour.
This critical boson contour can appear, for instance, 
on the interface of magnetic heterostructure~\cite{Zhang2021} 
or in 2D frustrated magnets~\cite{PhysRevResearch.2.033260}.
Particular interest lies in the situation where the radius of the boson contour 
can be  commensurate or incommensurate to the Fermi circle. 
For the commensurate cases, only finite Fermi points are connected by the boson contour. 
The crossover/transition to incommensurate cases bridges the infinite critical boson NFL 
with the conventional one described by the hotspot theory.
Moreover, one can consider the fermion-boson-coupled system in mixed dimensions,
namely, the dimension of the Fermi surface is incompatible with critical boson modes.
One intuitive example can be found in the 3D fractional quantum Hall system~\cite{Levin2009}.
Soft gauge bosons in 3D bulk can couple with the chiral Fermi level of the partons on the 2D surface~\cite{Levin2009},
which may trigger a NFL instability on the surface.
Beyond the scope of condensed matter physics, this general framework may 
apply to the meson-neutron coupling in neutron stars 
where the meson condensation forms a degenerate boson 
surface~\cite{Migdal}.\\

\noindent{\bf\small Data Availability}\\
{ \noindent The possible data that support the findings of this study
are available from the corresponding author (G.C.) upon request.}
\\

\noindent{{\bf References.}}
\bibliography{Ref.bib}

\begin{thebibliography}{10}
\expandafter\ifx\csname url\endcsname\relax
  \def\url#1{\texttt{#1}}\fi
\expandafter\ifx\csname urlprefix\endcsname\relax\def\urlprefix{URL }\fi
\providecommand{\bibinfo}[2]{#2}
\providecommand{\eprint}[2][]{\url{#2}}

\bibitem{Landau}
\bibinfo{author}{Landau, L.~D.}, \bibinfo{author}{Lifshitz, E.~M.} \&
  \bibinfo{author}{Pitaevskii, L.}
\newblock \emph{\bibinfo{title}{Statistical physics: theory of the condensed
  state}} (\bibinfo{publisher}{Butterworth-Heinemann}, \bibinfo{year}{1980}).

\bibitem{RMP2007}
\bibinfo{author}{L\"ohneysen, H.~v.}, \bibinfo{author}{Rosch, A.},
  \bibinfo{author}{Vojta, M.} \& \bibinfo{author}{W\"olfle, P.}
\newblock \bibinfo{title}{Fermi-liquid instabilities at magnetic quantum phase
  transitions}.
\newblock \emph{\bibinfo{journal}{Rev. Mod. Phys.}}
  \textbf{\bibinfo{volume}{79}}, \bibinfo{pages}{1015--1075}
  (\bibinfo{year}{2007}).

\bibitem{Hertz1976}
\bibinfo{author}{Hertz, J.~A.}
\newblock \bibinfo{title}{Quantum critical phenomena}.
\newblock \emph{\bibinfo{journal}{Phys. Rev. B}} \textbf{\bibinfo{volume}{14}},
  \bibinfo{pages}{1165--1184} (\bibinfo{year}{1976}).

\bibitem{Watanabe_2014}
\bibinfo{author}{Watanabe, H.} \& \bibinfo{author}{Vishwanath, A.}
\newblock \bibinfo{title}{{Criterion for stability of Goldstone modes and Fermi
  liquid behavior in a metal with broken symmetry}}.
\newblock \emph{\bibinfo{journal}{Proc. Natl. Acad. Sci. U.S.A.}}
  \textbf{\bibinfo{volume}{111}}, \bibinfo{pages}{16314--16318}
  (\bibinfo{year}{2014}).

\bibitem{SSLee2018}
\bibinfo{author}{Lee, S.-S.}
\newblock \bibinfo{title}{{Recent developments in non-Fermi liquid theory}}.
\newblock \emph{\bibinfo{journal}{Annu. Rev. Condens. Matter Phys.}}
  \textbf{\bibinfo{volume}{9}}, \bibinfo{pages}{227--244}
  (\bibinfo{year}{2018}).

\bibitem{SSLee2009}
\bibinfo{author}{Lee, S.-S.}
\newblock \bibinfo{title}{{Low-energy effective theory of Fermi surface coupled
  with U(1) gauge field in $2+1$ dimensions}}.
\newblock \emph{\bibinfo{journal}{Phys. Rev. B}} \textbf{\bibinfo{volume}{80}},
  \bibinfo{pages}{165102} (\bibinfo{year}{2009}).

\bibitem{Millis1993}
\bibinfo{author}{Millis, A.~J.}
\newblock \bibinfo{title}{Effect of a nonzero temperature on quantum critical
  points in itinerant fermion systems}.
\newblock \emph{\bibinfo{journal}{Phys. Rev. B}} \textbf{\bibinfo{volume}{48}},
  \bibinfo{pages}{7183--7196} (\bibinfo{year}{1993}).

\bibitem{Moriya1973}
\bibinfo{author}{Moriya, T.} \& \bibinfo{author}{Kawabata, A.}
\newblock \bibinfo{title}{{Effect of spin fluctuations on itinerant electron
  ferromagnetism}}.
\newblock \emph{\bibinfo{journal}{J. Phys. Soc. Jpn.}}
  \textbf{\bibinfo{volume}{34}}, \bibinfo{pages}{639--651}
  (\bibinfo{year}{1973}).

\bibitem{Moriya1973B}
\bibinfo{author}{Moriya, T.} \& \bibinfo{author}{Kawabata, A.}
\newblock \bibinfo{title}{{Effect of spin fluctuations on itinerant electron
  ferromagnetism. {II}}}.
\newblock \emph{\bibinfo{journal}{J. Phys. Soc. Jpn.}}
  \textbf{\bibinfo{volume}{35}}, \bibinfo{pages}{669--676}
  (\bibinfo{year}{1973}).

\bibitem{Moriya2012}
\bibinfo{author}{Moriya, T.}
\newblock \emph{\bibinfo{title}{Spin fluctuations in itinerant electron
  magnetism}} (\bibinfo{publisher}{Springer Science \& Business Media},
  \bibinfo{year}{2012}).

\bibitem{Chubukov2000}
\bibinfo{author}{Abanov, A.} \& \bibinfo{author}{Chubukov, A.~V.}
\newblock \bibinfo{title}{Spin-fermion model near the quantum critical point:
  One-loop renormalization group results}.
\newblock \emph{\bibinfo{journal}{Phys. Rev. Lett.}}
  \textbf{\bibinfo{volume}{84}}, \bibinfo{pages}{5608--5611}
  (\bibinfo{year}{2000}).

\bibitem{Chubukov2004}
\bibinfo{author}{Abanov, A.} \& \bibinfo{author}{Chubukov, A.}
\newblock \bibinfo{title}{Anomalous scaling at the quantum critical point in
  itinerant antiferromagnets}.
\newblock \emph{\bibinfo{journal}{Phys. Rev. Lett.}}
  \textbf{\bibinfo{volume}{93}}, \bibinfo{pages}{255702}
  (\bibinfo{year}{2004}).

\bibitem{Abrahams2012}
\bibinfo{author}{Abrahams, E.} \& \bibinfo{author}{Wolfle, P.}
\newblock \bibinfo{title}{Critical quasiparticle theory applied to heavy
  fermion metals near an antiferromagnetic quantum phase transition}.
\newblock \emph{\bibinfo{journal}{Proc. Natl. Acad. Sci. U.S.A.}}
  \textbf{\bibinfo{volume}{109}}, \bibinfo{pages}{3238--3242}
  (\bibinfo{year}{2012}).

\bibitem{SSLee2015}
\bibinfo{author}{Sur, S.} \& \bibinfo{author}{Lee, S.-S.}
\newblock \bibinfo{title}{Quasilocal strange metal}.
\newblock \emph{\bibinfo{journal}{Phys. Rev. B}} \textbf{\bibinfo{volume}{91}},
  \bibinfo{pages}{125136} (\bibinfo{year}{2015}).

\bibitem{SSLee2017}
\bibinfo{author}{Schlief, A.}, \bibinfo{author}{Lunts, P.} \&
  \bibinfo{author}{Lee, S.-S.}
\newblock \bibinfo{title}{{Exact critical exponents for the antiferromagnetic
  quantum critical metal in two dimensions}}.
\newblock \emph{\bibinfo{journal}{Phys. Rev. X}} \textbf{\bibinfo{volume}{7}},
  \bibinfo{pages}{021010} (\bibinfo{year}{2017}).

\bibitem{Sachdev2010B}
\bibinfo{author}{Metlitski, M.~A.} \& \bibinfo{author}{Sachdev, S.}
\newblock \bibinfo{title}{{Quantum phase transitions of metals in two spatial
  dimensions. II. Spin density wave order}}.
\newblock \emph{\bibinfo{journal}{Phys. Rev. B}} \textbf{\bibinfo{volume}{82}},
  \bibinfo{pages}{075128} (\bibinfo{year}{2010}).

\bibitem{Fradkin2001}
\bibinfo{author}{Oganesyan, V.}, \bibinfo{author}{Kivelson, S.~A.} \&
  \bibinfo{author}{Fradkin, E.}
\newblock \bibinfo{title}{{Quantum theory of a nematic Fermi fluid}}.
\newblock \emph{\bibinfo{journal}{Phys. Rev. B}} \textbf{\bibinfo{volume}{64}},
  \bibinfo{pages}{195109} (\bibinfo{year}{2001}).

\bibitem{Metzner2003}
\bibinfo{author}{Metzner, W.}, \bibinfo{author}{Rohe, D.} \&
  \bibinfo{author}{Andergassen, S.}
\newblock \bibinfo{title}{{Soft Fermi surfaces and breakdown of Fermi-liquid
  behavior}}.
\newblock \emph{\bibinfo{journal}{Phys. Rev. Lett.}}
  \textbf{\bibinfo{volume}{91}}, \bibinfo{pages}{066402}
  (\bibinfo{year}{2003}).

\bibitem{Fradkin2007}
\bibinfo{author}{Lawler, M.~J.} \& \bibinfo{author}{Fradkin, E.}
\newblock \bibinfo{title}{Local quantum criticality at the nematic quantum
  phase transition}.
\newblock \emph{\bibinfo{journal}{Phys. Rev. B}} \textbf{\bibinfo{volume}{75}},
  \bibinfo{pages}{033304} (\bibinfo{year}{2007}).

\bibitem{Sachdev2010}
\bibinfo{author}{Metlitski, M.~A.} \& \bibinfo{author}{Sachdev, S.}
\newblock \bibinfo{title}{{Quantum phase transitions of metals in two spatial
  dimensions. I. Ising-nematic order}}.
\newblock \emph{\bibinfo{journal}{Phys. Rev. B}} \textbf{\bibinfo{volume}{82}},
  \bibinfo{pages}{075127} (\bibinfo{year}{2010}).

\bibitem{SSLee2013}
\bibinfo{author}{Dalidovich, D.} \& \bibinfo{author}{Lee, S.-S.}
\newblock \bibinfo{title}{{Perturbative non-Fermi liquids from dimensional
  regularization}}.
\newblock \emph{\bibinfo{journal}{Phys. Rev. B}} \textbf{\bibinfo{volume}{88}},
  \bibinfo{pages}{245106} (\bibinfo{year}{2013}).

\bibitem{Chubukov2006}
\bibinfo{author}{Rech, J.}, \bibinfo{author}{P\'epin, C.} \&
  \bibinfo{author}{Chubukov, A.~V.}
\newblock \bibinfo{title}{{Quantum critical behavior in itinerant electron
  systems: Eliashberg theory and instability of a ferromagnetic quantum
  critical point}}.
\newblock \emph{\bibinfo{journal}{Phys. Rev. B}} \textbf{\bibinfo{volume}{74}},
  \bibinfo{pages}{195126} (\bibinfo{year}{2006}).

\bibitem{Chubukov2009}
\bibinfo{author}{Chubukov, A.~V.} \& \bibinfo{author}{Maslov, D.~L.}
\newblock \bibinfo{title}{{Spin conservation and Fermi liquid near a
  ferromagnetic quantum critical point}}.
\newblock \emph{\bibinfo{journal}{Phys. Rev. Lett.}}
  \textbf{\bibinfo{volume}{103}}, \bibinfo{pages}{216401}
  (\bibinfo{year}{2009}).

\bibitem{Polchinski1994}
\bibinfo{author}{Polchinski, J.}
\newblock \bibinfo{title}{Low-energy dynamics of the spinon-gauge system}.
\newblock \emph{\bibinfo{journal}{Nucl. Phys. B}}
  \textbf{\bibinfo{volume}{422}}, \bibinfo{pages}{617--633}
  (\bibinfo{year}{1994}).

\bibitem{Millis1994}
\bibinfo{author}{Altshuler, B.~L.}, \bibinfo{author}{Ioffe, L.~B.} \&
  \bibinfo{author}{Millis, A.~J.}
\newblock \bibinfo{title}{Low-energy properties of fermions with singular
  interactions}.
\newblock \emph{\bibinfo{journal}{Phys. Rev. B}} \textbf{\bibinfo{volume}{50}},
  \bibinfo{pages}{14048--14064} (\bibinfo{year}{1994}).

\bibitem{Kim1995}
\bibinfo{author}{Kim, Y.~B.}, \bibinfo{author}{Lee, P.~A.} \&
  \bibinfo{author}{Wen, X.-G.}
\newblock \bibinfo{title}{{Quantum Boltzmann equation of composite fermions
  interacting with a gauge field}}.
\newblock \emph{\bibinfo{journal}{Phys. Rev. B}} \textbf{\bibinfo{volume}{52}},
  \bibinfo{pages}{17275--17292} (\bibinfo{year}{1995}).

\bibitem{Nayak1994}
\bibinfo{author}{Nayak, C.} \& \bibinfo{author}{Wilczek, F.}
\newblock \bibinfo{title}{{Non-Fermi liquid fixed point in $2+1$ dimensions}}.
\newblock \emph{\bibinfo{journal}{Nucl. Phys. B}}
  \textbf{\bibinfo{volume}{417}}, \bibinfo{pages}{359--373}
  (\bibinfo{year}{1994}).

\bibitem{Senthil2010}
\bibinfo{author}{Mross, D.~F.}, \bibinfo{author}{McGreevy, J.},
  \bibinfo{author}{Liu, H.} \& \bibinfo{author}{Senthil, T.}
\newblock \bibinfo{title}{{Controlled expansion for certain non-Fermi-liquid
  metals}}.
\newblock \emph{\bibinfo{journal}{Phys. Rev. B}} \textbf{\bibinfo{volume}{82}},
  \bibinfo{pages}{045121} (\bibinfo{year}{2010}).

\bibitem{PRL1995}
\bibinfo{author}{Chakravarty, S.}, \bibinfo{author}{Norton, R.~E.} \&
  \bibinfo{author}{Sylju\aa{}sen, O.~F.}
\newblock \bibinfo{title}{{Transverse gauge interactions and the vanquished
  Fermi liquid}}.
\newblock \emph{\bibinfo{journal}{Phys. Rev. Lett.}}
  \textbf{\bibinfo{volume}{74}}, \bibinfo{pages}{1423--1426}
  (\bibinfo{year}{1995}).

\bibitem{Shankar2009}
\bibinfo{author}{Senthil, T.} \& \bibinfo{author}{Shankar, R.}
\newblock \bibinfo{title}{Fermi surfaces in general codimension and a new
  controlled nontrivial fixed point}.
\newblock \emph{\bibinfo{journal}{Phys. Rev. Lett.}}
  \textbf{\bibinfo{volume}{102}}, \bibinfo{pages}{046406}
  (\bibinfo{year}{2009}).

\bibitem{SSLee2015B}
\bibinfo{author}{Mandal, I.} \& \bibinfo{author}{Lee, S.-S.}
\newblock \bibinfo{title}{{Ultraviolet/infrared mixing in non-Fermi liquids}}.
\newblock \emph{\bibinfo{journal}{Phys. Rev. B}} \textbf{\bibinfo{volume}{92}},
  \bibinfo{pages}{035141} (\bibinfo{year}{2015}).

\bibitem{Giamarchi2003}
\bibinfo{author}{Giamarchi, T.}
\newblock \emph{\bibinfo{title}{Quantum physics in one dimension}}
  (\bibinfo{publisher}{Clarendon press}, \bibinfo{year}{2003}).

\bibitem{Das1999}
\bibinfo{author}{Das, D.} \& \bibinfo{author}{Doniach, S.}
\newblock \bibinfo{title}{{Existence of a Bose metal at $T=0$}}.
\newblock \emph{\bibinfo{journal}{Phys. Rev. B}} \textbf{\bibinfo{volume}{60}},
  \bibinfo{pages}{1261--1275} (\bibinfo{year}{1999}).

\bibitem{Phillips2003}
\bibinfo{author}{Phillips, P.}
\newblock \bibinfo{title}{{The elusive Bose metal}}.
\newblock \emph{\bibinfo{journal}{Science}} \textbf{\bibinfo{volume}{302}},
  \bibinfo{pages}{243--247} (\bibinfo{year}{2003}).

\bibitem{Paramekanti2002}
\bibinfo{author}{Paramekanti, A.}, \bibinfo{author}{Balents, L.} \&
  \bibinfo{author}{Fisher, M. P.~A.}
\newblock \bibinfo{title}{{Ring exchange, the exciton Bose liquid, and
  bosonization in two dimensions}}.
\newblock \emph{\bibinfo{journal}{Phys. Rev. B}} \textbf{\bibinfo{volume}{66}},
  \bibinfo{pages}{054526} (\bibinfo{year}{2002}).

\bibitem{Motrunich2007}
\bibinfo{author}{Motrunich, O.~I.} \& \bibinfo{author}{Fisher, M. P.~A.}
\newblock \bibinfo{title}{{$d$-wave correlated critical Bose liquids in two
  dimensions}}.
\newblock \emph{\bibinfo{journal}{Phys. Rev. B}} \textbf{\bibinfo{volume}{75}},
  \bibinfo{pages}{235116} (\bibinfo{year}{2007}).

\bibitem{Sheng2009}
\bibinfo{author}{Sheng, D.~N.}, \bibinfo{author}{Motrunich, O.~I.} \&
  \bibinfo{author}{Fisher, M. P.~A.}
\newblock \bibinfo{title}{{Spin Bose-metal phase in a spin-$\frac{1}{2}$ model
  with ring exchange on a two-leg triangular strip}}.
\newblock \emph{\bibinfo{journal}{Phys. Rev. B}} \textbf{\bibinfo{volume}{79}},
  \bibinfo{pages}{205112} (\bibinfo{year}{2009}).

\bibitem{Han2021}
\bibinfo{author}{Han, S.} \& \bibinfo{author}{Kim, Y.~B.}
\newblock \bibinfo{title}{{Non-Fermi liquid induced by Bose metal with
  protected subsystem symmetries}}.
\newblock \emph{\bibinfo{journal}{Phys. Rev. B}}
  \textbf{\bibinfo{volume}{106}}, \bibinfo{pages}{L081106}
  (\bibinfo{year}{2022}).

\bibitem{Sur2019}
\bibinfo{author}{Sur, S.} \& \bibinfo{author}{Yang, K.}
\newblock \bibinfo{title}{Metallic state in bosonic systems with continuously
  degenerate dispersion minima}.
\newblock \emph{\bibinfo{journal}{Phys. Rev. B}}
  \textbf{\bibinfo{volume}{100}}, \bibinfo{pages}{024519}
  (\bibinfo{year}{2019}).

\bibitem{Lake2021}
\bibinfo{author}{Lake, E.}, \bibinfo{author}{Senthil, T.} \&
  \bibinfo{author}{Vishwanath, A.}
\newblock \bibinfo{title}{{Bose-Luttinger liquids}}.
\newblock \emph{\bibinfo{journal}{Phys. Rev. B}}
  \textbf{\bibinfo{volume}{104}}, \bibinfo{pages}{014517}
  (\bibinfo{year}{2021}).

\bibitem{Pfleiderer2001}
\bibinfo{author}{Pfleiderer, C.}, \bibinfo{author}{Julian, S.~R.} \&
  \bibinfo{author}{Lonzarich, G.~G.}
\newblock \bibinfo{title}{{Non-Fermi-liquid nature of the normal state of
  itinerant-electron ferromagnets}}.
\newblock \emph{\bibinfo{journal}{Nature}} \textbf{\bibinfo{volume}{414}},
  \bibinfo{pages}{427--430} (\bibinfo{year}{2001}).

\bibitem{Pfleiderer2004}
\bibinfo{author}{Pfleiderer, C.} \emph{et~al.}
\newblock \bibinfo{title}{{Partial order in the non-Fermi-liquid phase of
  {MnSi}}}.
\newblock \emph{\bibinfo{journal}{Nature}} \textbf{\bibinfo{volume}{427}},
  \bibinfo{pages}{227--231} (\bibinfo{year}{2004}).

\bibitem{Tsvelik2021}
\bibinfo{author}{Lajer, M.}, \bibinfo{author}{Konik, R.~M.},
  \bibinfo{author}{Pisarski, R.~D.} \& \bibinfo{author}{Tsvelik, A.~M.}
\newblock \bibinfo{title}{{When cold, dense quarks in $1+1$ and $3+1$
  dimensions are not a Fermi liquid}}.
\newblock \emph{\bibinfo{journal}{Phys. Rev. D}}
  \textbf{\bibinfo{volume}{105}}, \bibinfo{pages}{054035}
  (\bibinfo{year}{2022}).

\bibitem{Ku2021}
\bibinfo{author}{Hegg, A.}, \bibinfo{author}{Hou, J.} \& \bibinfo{author}{Ku,
  W.}
\newblock \bibinfo{title}{{Geometric frustration produces long-sought Bose
  metal phase of quantum matter}}.
\newblock \emph{\bibinfo{journal}{Proc. Natl. Acad. Sci. U.S.A.}}
  \textbf{\bibinfo{volume}{118}} (\bibinfo{year}{2021}).

\bibitem{Zhang2022}
\bibinfo{author}{Pan, Z.} \& \bibinfo{author}{Zhang, X.-T.}
\newblock \bibinfo{title}{{Infinite critical boson induced non-Fermi liquid in
  $d=3-\epsilon$ dimensions}}.
\newblock \eprint{Preprint at https://arxiv.org/abs/2205.03818 (2022)}.

\bibitem{Brazovskii1975}
\bibinfo{author}{Brazovskii, S.~A.}
\newblock \bibinfo{title}{Phase transition of an isotropic system to an
  inhomogeneous state}.
\newblock \emph{\bibinfo{journal}{Sov. Phys. JETP}}
  \textbf{\bibinfo{volume}{41}}, \bibinfo{pages}{85--89}
  (\bibinfo{year}{1975}).

\bibitem{Schmalian2004}
\bibinfo{author}{Schmalian, J.} \& \bibinfo{author}{Turlakov, M.}
\newblock \bibinfo{title}{Quantum phase transitions of magnetic rotons}.
\newblock \emph{\bibinfo{journal}{Phys. Rev. Lett.}}
  \textbf{\bibinfo{volume}{93}}, \bibinfo{pages}{036405}
  (\bibinfo{year}{2004}).

\bibitem{Vojta2001}
\bibinfo{author}{Vojta, T.} \& \bibinfo{author}{Sknepnek, R.}
\newblock \bibinfo{title}{Quantum phase transition of itinerant helimagnets}.
\newblock \emph{\bibinfo{journal}{Phys. Rev. B}} \textbf{\bibinfo{volume}{64}},
  \bibinfo{pages}{052404} (\bibinfo{year}{2001}).

\bibitem{Janoschek2013}
\bibinfo{author}{Janoschek, M.} \emph{et~al.}
\newblock \bibinfo{title}{{Fluctuation-induced first-order phase transition in
  Dzyaloshinskii-Moriya helimagnets}}.
\newblock \emph{\bibinfo{journal}{Phys. Rev. B}} \textbf{\bibinfo{volume}{87}},
  \bibinfo{pages}{134407} (\bibinfo{year}{2013}).

\bibitem{nagaosa1999}
\bibinfo{author}{Nagaosa, N.}
\newblock \emph{\bibinfo{title}{Quantum field theory in strongly correlated
  electronic systems}} (\bibinfo{publisher}{Springer Science \& Business
  Media}, \bibinfo{year}{1999}).

\bibitem{BK2006}
\bibinfo{author}{Belitz, D.}, \bibinfo{author}{Kirkpatrick, T.~R.} \&
  \bibinfo{author}{Rosch, A.}
\newblock \bibinfo{title}{Theory of helimagnons in itinerant quantum systems}.
\newblock \emph{\bibinfo{journal}{Phys. Rev. B}} \textbf{\bibinfo{volume}{73}},
  \bibinfo{pages}{054431} (\bibinfo{year}{2006}).

\bibitem{BK2005}
\bibinfo{author}{Kirkpatrick, T.~R.} \& \bibinfo{author}{Belitz, D.}
\newblock \bibinfo{title}{{Nonanalytic corrections to Fermi-liquid behavior in
  helimagnets}}.
\newblock \emph{\bibinfo{journal}{Phys. Rev. B}} \textbf{\bibinfo{volume}{72}},
  \bibinfo{pages}{180402} (\bibinfo{year}{2005}).

\bibitem{BK2006B}
\bibinfo{author}{Belitz, D.}, \bibinfo{author}{Kirkpatrick, T.~R.} \&
  \bibinfo{author}{Rosch, A.}
\newblock \bibinfo{title}{{Theory of helimagnons in itinerant quantum systems.
  II. Nonanalytic corrections to Fermi-liquid behavior}}.
\newblock \emph{\bibinfo{journal}{Phys. Rev. B}} \textbf{\bibinfo{volume}{74}},
  \bibinfo{pages}{024409} (\bibinfo{year}{2006}).

\bibitem{Zhang2021}
\bibinfo{author}{Zhang, X.-T.} \& \bibinfo{author}{Chen, G.}
\newblock \bibinfo{title}{{Infinite critical boson non-Fermi liquid on
  heterostructure interfaces}}.
\newblock \eprint{Preprint at https://arxiv.org/abs/2109.06594 (2021)}.

\bibitem{PhysRevResearch.2.033260}
\bibinfo{author}{Liu, J.~Q.}, \bibinfo{author}{Li, F.-Y.},
  \bibinfo{author}{Chen, G.} \& \bibinfo{author}{Wang, Z.}
\newblock \bibinfo{title}{{Featureless quantum paramagnet with frustrated
  criticality and competing spiral magnetism on spin-1 honeycomb lattice
  magnet}}.
\newblock \emph{\bibinfo{journal}{Phys. Rev. Research}}
  \textbf{\bibinfo{volume}{2}}, \bibinfo{pages}{033260} (\bibinfo{year}{2020}).

\bibitem{Levin2009}
\bibinfo{author}{Levin, M.} \& \bibinfo{author}{Fisher, M. P.~A.}
\newblock \bibinfo{title}{{Gapless layered three-dimensional fractional quantum
  Hall states}}.
\newblock \emph{\bibinfo{journal}{Phys. Rev. B}} \textbf{\bibinfo{volume}{79}},
  \bibinfo{pages}{235315} (\bibinfo{year}{2009}).

\bibitem{Migdal}
\bibinfo{author}{Migdal, A.}, \bibinfo{author}{Saperstein, E.},
  \bibinfo{author}{Troitsky, M.} \& \bibinfo{author}{Voskresensky, D.}
\newblock \bibinfo{title}{Pion degrees of freedom in nuclear matter}.
\newblock \emph{\bibinfo{journal}{Phys. Rep.}} \textbf{\bibinfo{volume}{192}},
  \bibinfo{pages}{179--437} (\bibinfo{year}{1990}).

\end{thebibliography}
\vspace{0.5cm}

\noindent{\emph{\bf Acknowledgments}}\\
We thank Yonghao Gao, Sungsik Lee, Michael Hermele, 
Arun Paramekanti, and Leo Radzihovsky for discussion. 
This work is supported by the Ministry of Science and Technology of China 
with Grants No.~2021YFA1400300, by the National Science Foundation of China with Grant No.~92065203,
and by the Research Grants Council of Hong Kong 
with General Research Fund Grant No.~17306520.
\\

\noindent{\emph{\bf Contributions}}\\
{\footnotesize \noindent{G.C.} designed and supervised this project. 
X.T.Z. and G.C. performed the calculation and wrote the manuscript.
}
\\

\noindent {\bf Additional Information}\\
{\footnotesize \noindent {\bf Reprints and permissions}
information is available online at www.nature.com/reprints.}\\

{\footnotesize \noindent
Correspondence and requests for materials should be addressed to
G.C. (gangchen@hku.hk).}\\

{\footnotesize \noindent \noindent {\bf Competing Interests}
The authors declare no competing financial or non-financial interests.}

\begin{widetext}

\newpage
\begin{figure}[htbp] 
	\centering
	\includegraphics[width=0.65\linewidth]{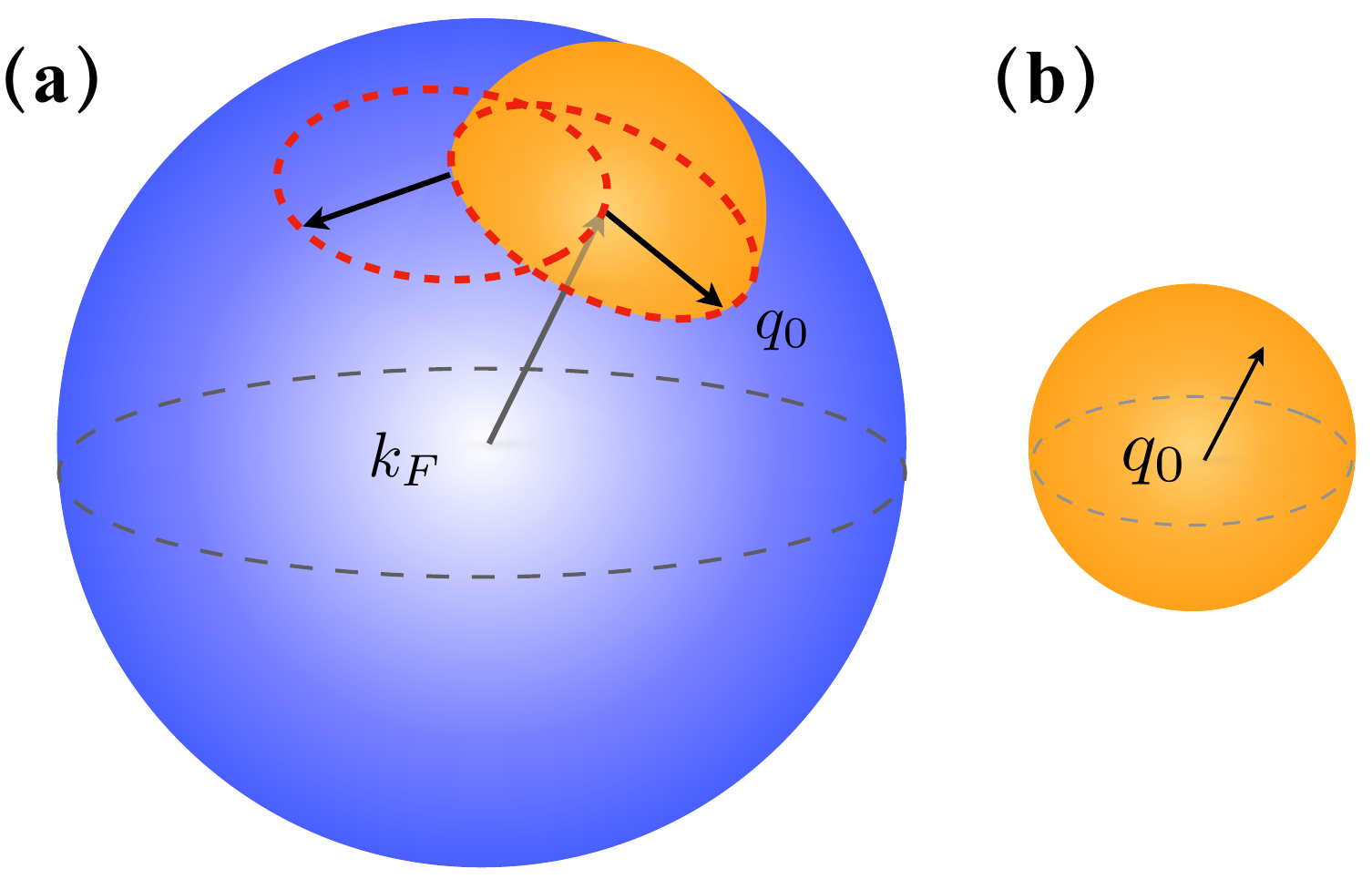}\\
\raggedright{Fig.~1: \small{
{\bf The coupling between 
	Fermi surface and critical boson surface.}
	(a) The blue sphere is the 
	Fermi surface of the conduction electrons.
	The orange surface represents
	the critical boson surface.
	$k_{\rm F}$ is the Fermi momentum, 
	and $q_0$ is the radius of the boson surface.
	The Yukawa coupling connects fermions 
	with a ring	of fermion modes on the Fermi surface.
	(b) The critical boson sphere. 
	}}
\end{figure}

\newpage
\begin{figure}[htbp] 
	\centering
	\includegraphics[width=0.65\linewidth]{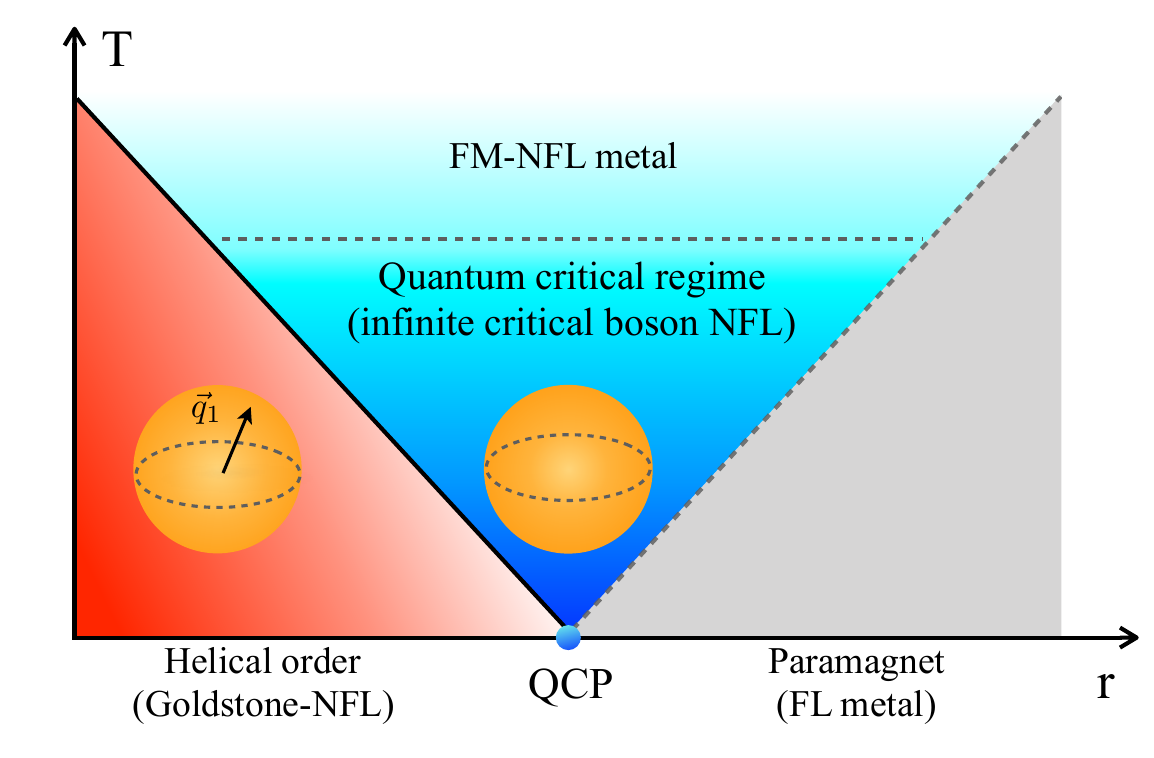}\\
\raggedright{Fig.~2: \small{
{\bf The global phase diagram.}
	The left corner is a helimagnet and a Goldstone-NFL where the NFL is induced by the gapless Goldstone boson. 
	The right corner is a paramagnet and Fermi liquid metal. The central region is the quantum critical regime where  
    a distinct NFL with infinite critical bosons on the boson surface is realized. 
    As the temperature rises to the point where the thermal fluctuation submerges the boson surface, 
    the system experiences a crossover to a FM criticality-like NFL behavior. 
    The solid (dashed) line refers to the phase transition (thermal crossover).
     	}}
\end{figure}

\newpage
\begin{figure}[htbp]  
	\centering
	\includegraphics[width=0.65\linewidth]{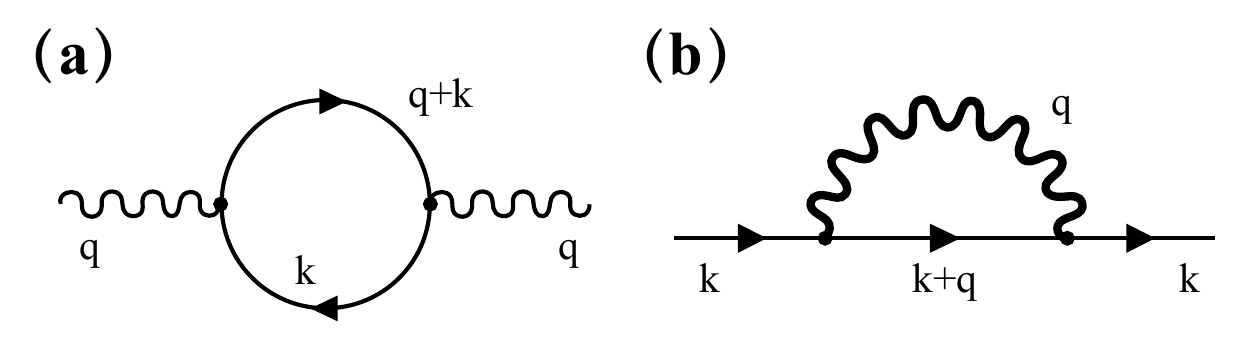}\\
	
\raggedright{Fig.~3: \small{
{\bf Renormalized bubble diagrams.}
	(a) The fermion bubble induced boson dynamics.  
	(b) The renormalized fermion propagator from the renormalized 
	boson correlator. The light and bold curly line represent the 
	bare and renormalized boson correlators, respectively. 
	}}
\end{figure} 

\end{widetext}

\end{document}